\documentclass[a4paper]{article}

\usepackage[margin=1.5in]{geometry}
\usepackage{booktabs} 

\usepackage[ruled]{algorithm2e} 
\usepackage{tikz} 
\usetikzlibrary{shapes.geometric, arrows}
\usepackage[colorinlistoftodos]{todonotes} 
\usepackage{authblk}
\usepackage{natbib}

\begin{document}

\title{A Vision for Health Informatics: \\ Introducing the SKED Framework\\
\Large An Extensible Architecture for Scientific Knowledge Extraction from Data}

\author[1]{Elizabeth D. Trippe}

\author[2]{Jacob B. Aguilar}

\author[1]{Yi H. Yan}

\author[3]{Mustafa V. Nural}

\author[4]{Jessica A. Brady}

\author[3]{Mehdi Assefi}

\author[3]{Saeid Safaei}
 
\author[3]{Mehdi Allahyari}

\author[3]{Seyedamin Pouriyeh}

\author[5]{Mary R. Galinski}

\author[1]{Jessica C. Kissinger}
 
\author[1,2,3]{Juan B. Gutierrez \footnote{Corresponding author: jgutierr@uga.edu}}

  \affil[1]{Institute of Bioinformatics, University of Georgia}
  \affil[2]{Department of Mathematics, University of Georgia}
  \affil[3]{Department of Computer Science, University of Georgia}
  \affil[4]{School of Engineering, University of Georgia}
  \affil[5]{Emory Vaccine Center, Emory University}

\maketitle

\begin{abstract}
The goals of the Triple Aim of health care and the goals of P4 medicine outline objectives that require a significant health informatics component. However, the goals do not provide specifications about how all of the new individual patient data will be combined in meaningful ways and with data from other sources, like epidemiological data, to promote the health of individuals and society. We seem to have more data than ever before but few resources and means to use it efficiently. We need a general, extensible solution that integrates and homogenizes data of disparate origin, incompatible formats, and multiple spatial and temporal scales. To address this problem, we introduce the Scientific Knowledge Extraction from Data (SKED) architecture, as a technology-agnostic framework to minimize the overhead of data integration, permit reuse of analytical pipelines, and guarantee reproducible quantitative results. The SKED architecture consists of a Resource Allocation Service to locate resources, and the definition of data primitives to simplify and harmonize data. SKED allows automated knowledge discovery and provides a platform for the realization of the major goals of modern health care.

\end{abstract}

\section{Introduction}
\begin{figure}[ht!]
\centering
\includegraphics[width=\linewidth]{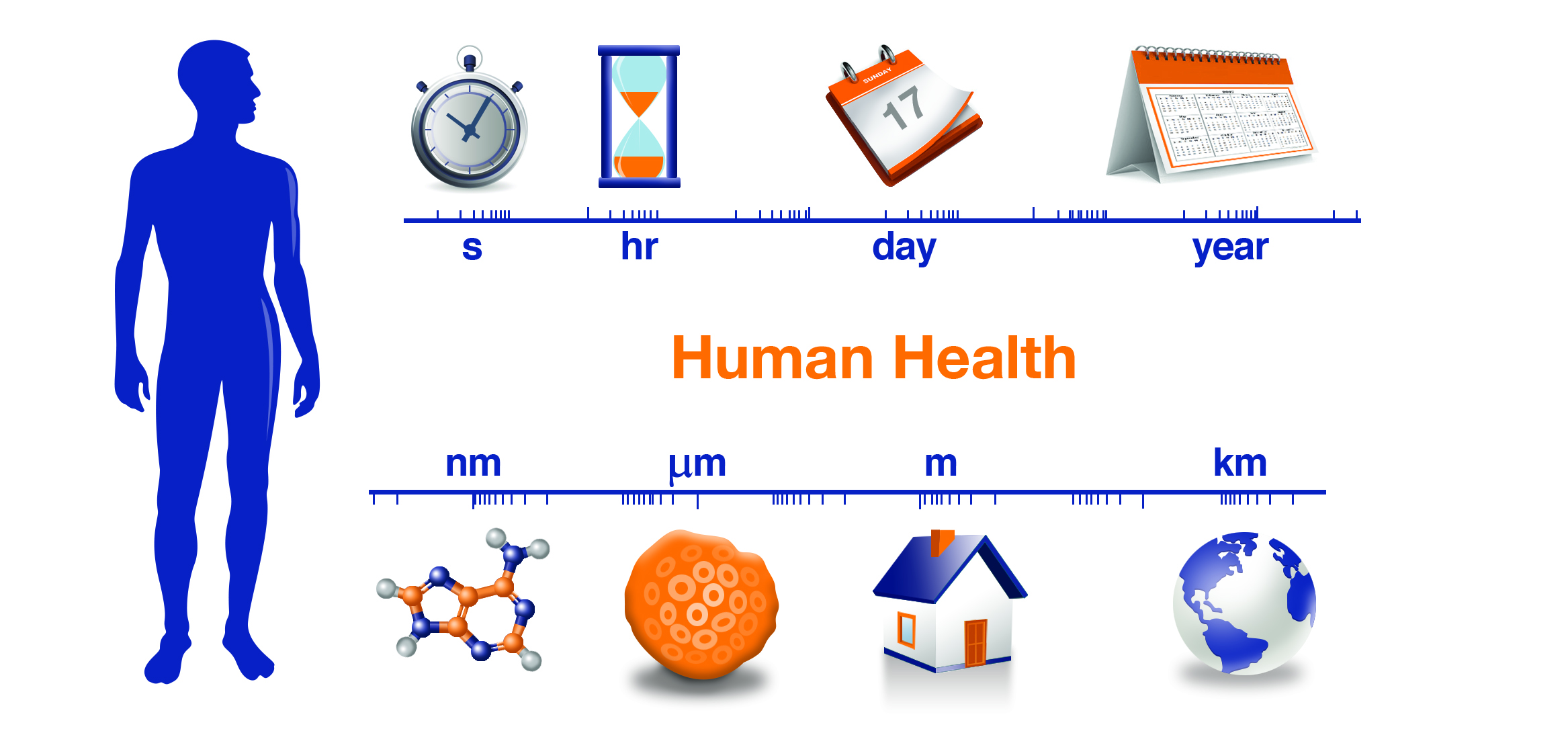}
 		\caption{Systems medicine incorporates patient data from multiple spatial and time scales  }
 		\label{fig:timeSpaceScale}
\end{figure}

Health informatics is a significant underlying component of the Triple Aim of health care which has goals of simultaneously improving the patient experience, improving the health of populations and reducing per capita costs  \cite{berwick2008triple}. As health care informatics begins to incorporate more P4 (predictive, preventative, personalized, participatory) systems medicine approaches, and to include patient measurements that have traditionally been used in research (genetic profiles, and other multi-omic technologies)  \cite{hood2012revolutionizing}, health informatics must integrate large heterogeneous datasets that cross temporal and spatial scales (see Figure  \ref{fig:timeSpaceScale}), to accomplish the goals of the Triple Aim. Most efforts so far have focused on creating detailed, workable solutions to manage these datasets in isolation but few have focused on their reconciliation. The magnitude of the problem is described in Figure \ref{fig:timeSpaceScale}. Molecular, cellular, clinical, environmental and epidemiological data have all been gathered in vast quantities to describe both individual patients and to characterize diseases, but this data has not resulted in significant improvements to individual patient care or reduced care costs. Currently there is no robust, scalable method to incorporate clinical information and other multi-omic datasets for routine patient care.  To address the informatics problems underlying P4 systems medicine and the Triple Aim of health care, we introduce the Scientific Knowledge Extraction from Data (SKED) architecture, a technology-agnostic framework to minimize the overhead of data integration, facilitate the reuse of analytical pipelines, and guarantee of reproducibility of quantitative results.

This manuscript is organized as follows. In Section 2, the SKED architecture will be introduced and described.  In Section 3, we will discuss how SKED enables automated knowledge generation.  In Section 4, we describe how we are using SKED in systems biology research and our success with multi-omic and clinical data integration. Last, in Sections 5 and 6, we outline our vision for expanding the use of the SKED framework in systems medicine. 

\section{Introducing the SKED Architecture}
\subsection{Data Primitives to Harmonize and Unify Data}
\begin{figure*}[ht!]
\begingroup%
  \makeatletter%
  \providecommand\color[2][]{%
    \errmessage{(Inkscape) Color is used for the text in Inkscape, but the package 'color.sty' is not loaded}%
    \renewcommand\color[2][]{}%
  }%
  \providecommand\transparent[1]{%
    \errmessage{(Inkscape) Transparency is used (non-zero) for the text in Inkscape, but the package 'transparent.sty' is not loaded}%
    \renewcommand\transparent[1]{}%
  }%
  \providecommand\rotatebox[2]{#2}%
  \ifx\svgwidth\undefined%
    \setlength{\unitlength}{\textwidth}
    \ifx\svgscale\undefined%
      \relax%
    \else%
      \setlength{\unitlength}{\unitlength * \real{\svgscale}}%
    \fi%
  \else%
    \setlength{\unitlength}{\svgwidth}%
  \fi%
  \global\let\svgwidth\undefined%
  \global\let\svgscale\undefined%
  \makeatother%
  \begin{picture}(1,0.41337026)%
    \put(0,0){\includegraphics[width=\unitlength,page=1]{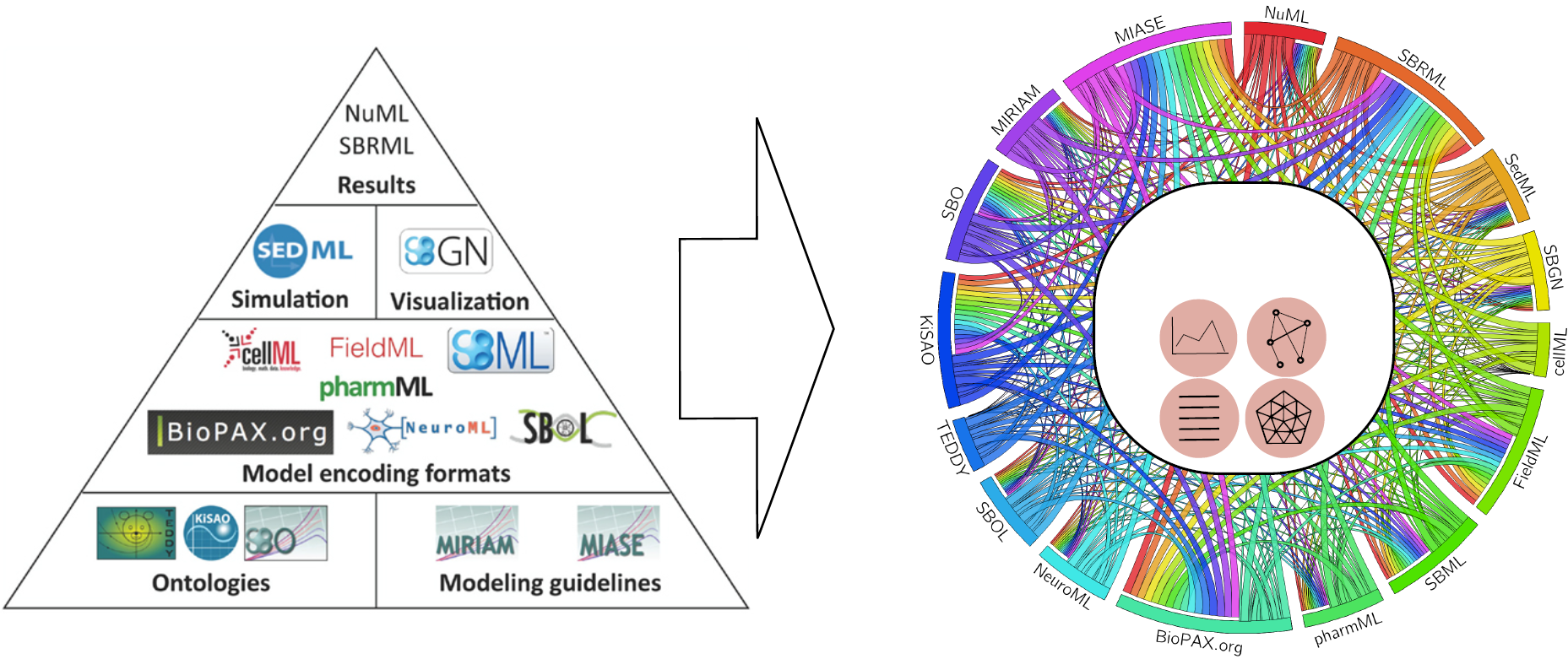}}%
    \put(0.12,0.42){\color[rgb]{0,0,0}\makebox(0,0)[lt]{\begin{minipage}{0.3\unitlength}\raggedright \textbf{\large State of the Art} \end{minipage}}}%
    \put(0.57,0.42){\color[rgb]{0,0,0}\makebox(0,0)[lt]{\begin{minipage}{0.11179428\unitlength}\raggedright \textbf{\large SKED} \end{minipage}}}%
    \put(0.73,0.29){\color[rgb]{0,0,0}\makebox(0,0)[lt]{\begin{minipage}{0.1\unitlength}\raggedright \centering \small \textbf{ Data \\ Primitives} \end{minipage}}}%
  \end{picture}%
\endgroup%

\caption{ \label{fig:StdstoDP} Data Primitives simplify how standards talk with each other (pyramid modified from Drager and Palsson, 2014 \cite{drager2014improving})  }%
\end{figure*}
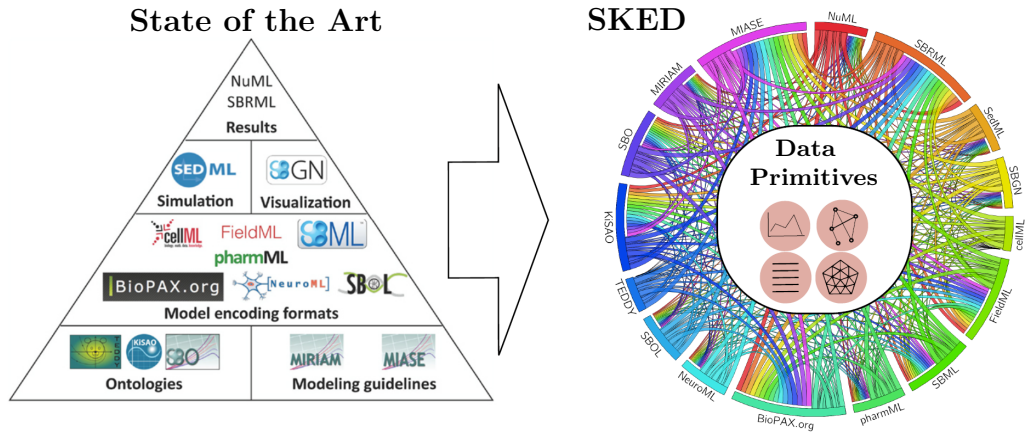

All data used in SKED conform to a reduced set of ``\textit{data primitives}'', \textit{i.e.} atomic units of data representation \textit{e.g.} time series, text, graphs and polygonal meshes. SKED ingests data from multiple repositories (parsers must be designed for each format), and transforms  them into data primitives.  Data primitives are independent of the underlying storage strategy. Current standards and ontologies function like puzzle pieces, which allow connections only to a reduced set of elements (as depicted in the left side of Figure \ref{fig:StdstoDP}.  Using data primitives, however, enables us to use data like LEGO\textregistered\space  building blocks, in which communication can occur between any two standards or combination of standards.   

Even though it has been stated that there is no ``one-size-fits-all'' standard for biological research  \cite{drager2014improving}, we believe that our efforts are fundamentally different and that many previous standardization efforts may be combined through the use of data primitives. This homogenization of data has deep implications for health informatics: Data primitives make it possible to seamlessly cross organizational domains in order to consolidate heterogeneous records. 

\subsection{The SKED Resource Allocation Service Locates Digital Resources}
The commoditization of high-performance and cloud computing has made machine learning algorithms broadly available for applications in large data sets at a cost level that permits mass adoption. Examples of cloud computing platforms are Amazon Web Services (AWS), Microsoft Azure (MA), and Google Cloud Platform (GCP), among others. These solutions offer similar functionality with different (and incompatible) proprietary implementations that also leverage open source tools.  

Solutions deployed in one platform could be, in principle, be used by any number of clients, e.g. machine learning pipelines to summarize patient records. But these tools rely on the specificities of the platforms in which they were developed, thus interested parties would need to adapt existing information systems or implement new ones to utilize these assets.  

With SKED, communication across systems is standardized by the use of data primitives. Data or computational assets would receive queries expressed in data primitives, and produce results only in data primitives. Hence, existing systems would only need a wrapper for data ingestion and data export. This paradigm trivializes the differences between cloud computing providers, in an analogous way in which video standards allow the same television set to be connected to many cable providers. 

\begin{figure}[ht!]

\centering
\includegraphics[width=0.7\linewidth]{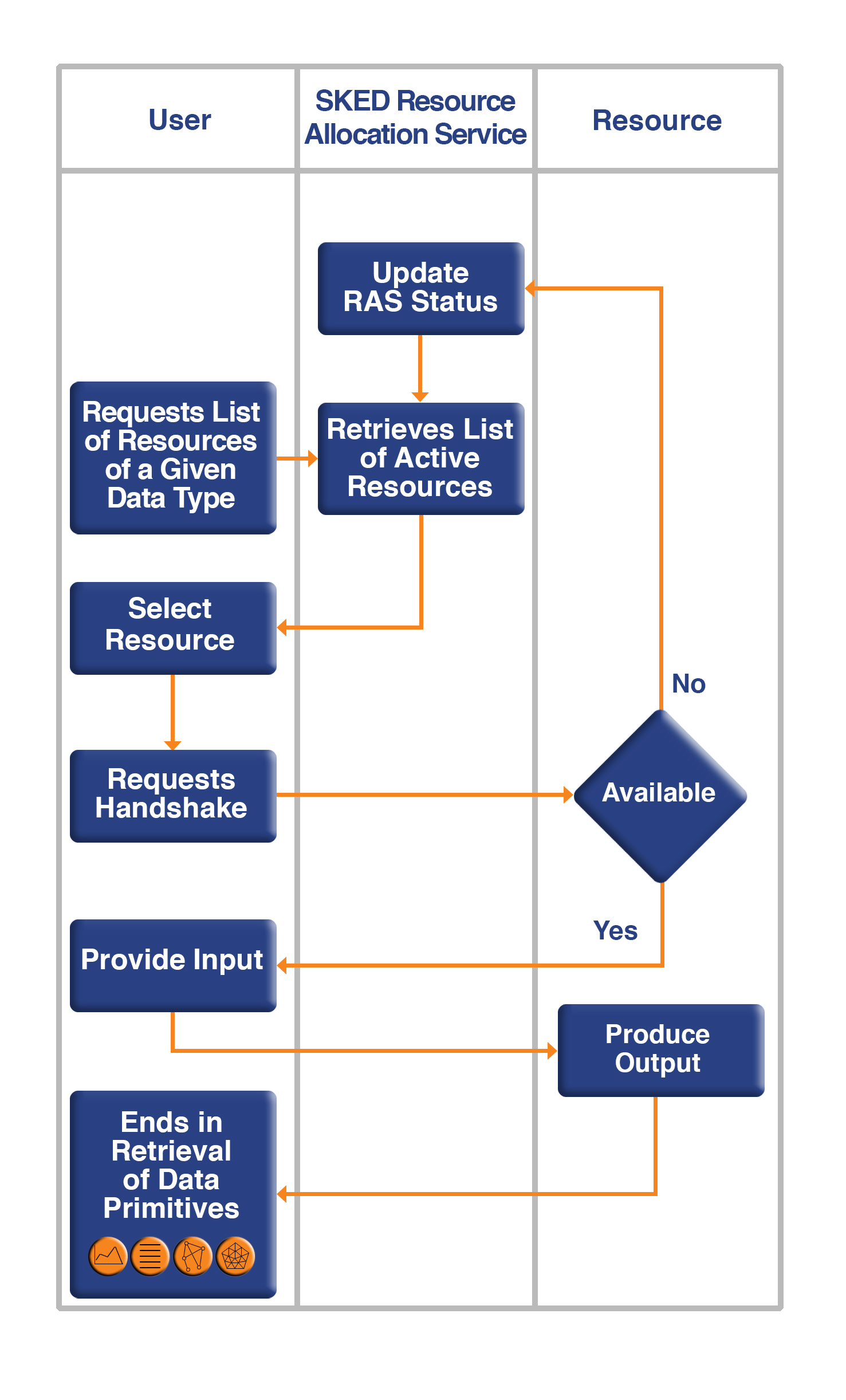}
\caption{ \label{fig:RAS} RAS allows researchers to query data, algorithms and pipelines}%

\end{figure}

However, the standardization of communication across systems requires a centralized resource that knows what assets are available at a given point in time, and with what capacity (e.g. bandwidth, FLOPS). This Resource Allocation Service (RAS) is implemented as a server set that contains a database of all known resources, e.g. datasets, pipelines, etc. RAS locates and allocates those requested resources. For each resource in the database there should be information about the address, available capacity, and share-ability. For each request, a RAS instance checks the validity of the request. This can be simply done by checking if there is a record associated to requested resource in the RAS database. The next step is checking the availability of the resource. 
If the resource is available, the RAS instance allocates the asset to the process and updates the database accordingly. The process gets access to the resource by receiving the resource address from the RAS instance. Resources can be released by the RAS server after a deadline or by sending a release message from the process to the server when the process is done with it. RAS updates the resource database whenever a resource is released. This process is depicted in Figure \ref{fig:RAS}. 
 
\section{SKED for Automated Knowledge Generation}
All modeling formalisms in SKED (causal models, generative models, discriminative models, finite state machines, categorical grammars, dynamical systems, logical systems graphical models, etc.)  accept as inputs only data primitives, and produce as outputs only data primitives. Thus, the overhead of reconfiguring these tools to study new problems is minimized and can be automated. 

\begin{figure}[ht!]

\centering
\includegraphics[width=0.7\linewidth]{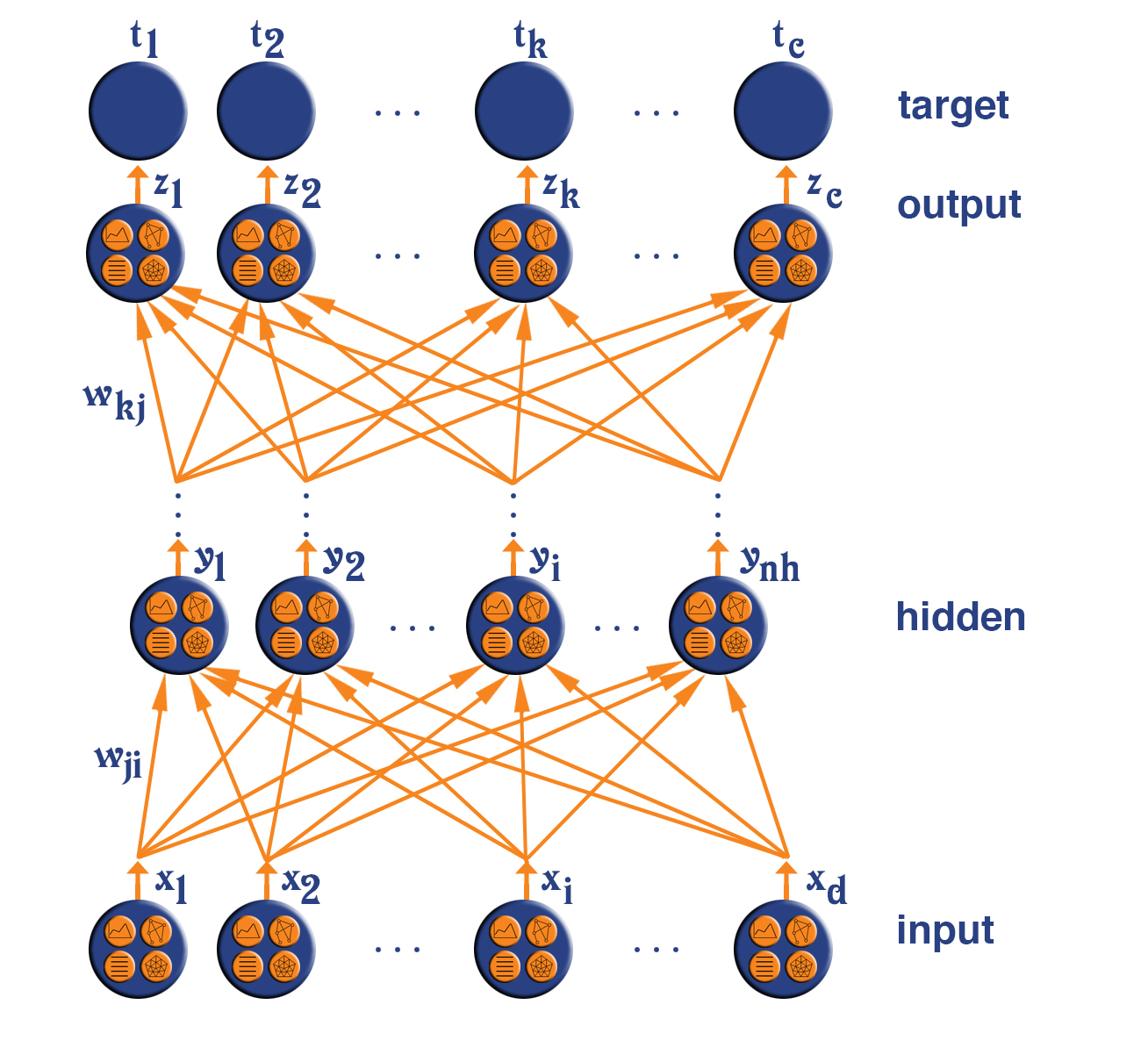}
\caption{\label{fig:multiLayAnalysis} Automated knowledge generation from SKED}

\end{figure}

Since RAS provides information about resources with specific inputs and outputs, an automated agent could concatenate existing resources (i.e. analysis pipelines, data) to produce a desired outcome. This process could produce both novel configurations to analyze data, and computational verification of results through consensus via the implementation of ensembles of methods. Figure  \ref{fig:multiLayAnalysis} describes automated knowledge generation using SKED.

The final targets could be classification into diseased or healthy states, or the prediction of disease progression. Building multi-scale models is thus made feasible using SKED, whether this be models of epidemiology about the spread of disease or whole-patient models (patient avatars) to predict the course of a disease. 

\section{SKED in Research}
We implemented the SKED framework to study the pathogenesis of malaria using multi-omic data (transcriptomics, proteomics, metabolomics, lipidomics), immunological data (flow cytometry, cytokine ELISAs), and clinical measurements (doctor assessments, and physio-telemetry), as part of the Malaria Host-Pathogen Interaction Center (MaHPIC). We investigated the host-pathogen interactions between non-human primates hosts and \textit{Plasmodium} parasites as models for human malarial infections.  

 We were able to combine high frequency telemetry signals (ex. ECG) with other measurements taken  over the course of an infection \cite{joyner2016plasmodium}, for example metabolomics (daily),  transcriptomics and immune response data (various times throughout the infection). Using data primitives allowed us to easily perform different types of meta-dimensional analysis as described by \citet{ritchie2015methods}, including concatenation-based analysis, where multiple data types are combined before analysis.  

One of the most powerful aspects of SKED was the ability to harmonize data over multiple time scales and multiple spatial scales and we envision this aspect to become even more important as additional real-time, continuous data measures (as could easily by gathered by e.g. a cell phone sensor) become available.


\section{SKED Enables P4 Medicine}

 Because the SKED framework provides a general, scalable solution to the problems of data integration and data harmonization across multiple time and spatial scales, patient treatments may be made more predictive as powerful algorithms that are able to identify the most important biomarkers for a disease are found. Algorithms designed for one type of data may be effortlessly repurposed for use on another data type. Concatenation-based analysis thus becomes more feasible and could allow for acceleration of biomarker discovery, since multiple-omic datasets may be combined in analysis \cite{ritchie2015methods}.
 
Chronic diseases (diabetes, cardiovascular disease, etc.) are now a major cause of mortality  in many countries; thus, biomarker discovery for early detection and intervention \cite{sagner2016p4} is a pressing need. SKED provides a workable solution to combine the complicated multi-omic data sets that must be gathered from many people in order to determine the most significant molecular predictors for these diseases.  

As predictions about the onset of chronic disease improve, these accurate predictions could enable earlier, preventative treatments to be undertaken.  The data integration capabilities of SKED establish a foundation for the use of more personalized medicine, as personalized medicine begins to make more use of genomic and other large-scale datasets to describe a patient.  As the ``individualome'' of each patient is created and becomes more complicated, patients could be able to have a more active part in managing their own health and outcomes \cite{shameer2017translational}. Patients will thus be better able to manage their own health and have a more active role in preventing the chronic disease they may be most susceptible to. 

\section{SKED enables the Triple Aim of Health Care}

Because the SKED framework solves many problems associated with data integration and harmonization at multiple levels in health information analysis, it is aligned with achieving the goals of the Triple Aim in health informatics  \cite{berwick2008triple}. \citet{whittington2015pursuing}  identify three principles that successfully guided organizations working on the implementation of the Triple Aim. 

The first guiding principle was establishing a foundation for population management to determine which populations (i.e. elderly, low-income, etc.) will be the focus of an intervention.  A system integrator (e.g. a local or state health department) gathers resources and coordinates work in this step.  The system integrator is also responsible for iterative improvements and testing to determine when and how the most short- and long-term progress has been made.  Such analysis can be done easily and effectively on the kinds of heterogeneous data that describe health outcomes using SKED.  SKED allows algorithms used in one context to be extended to others so that the most advanced up-to-date methods may be applied to any dataset to determine the effectiveness of an intervention.

The second guiding principle was to effectively manage services at scale. The SKED framework allows for the analysis of all types of data (epidemiological, clinical, etc.) at different scales.  Automated analysis with SKED could allow the most important services and their beneficial effects to be identified and subsequently implemented.  The results of implementing different health services at different scales may be studied and the most effective overall plans could be enabled through the use of SKED.

Last,  \citet{whittington2015pursuing} identified the need for a learning system to determine which measures have had the most effect. The authors propose that cycles of iterative testing are needed to investigate the performance of different interventions and treatments in populations and individuals.  Using data primitives in SKED can make such analyses more accurate and consistent. The RAS will simplify finding analysis pipelines and data for comparison.  For example, having data stored as data primitives could enable a public health official to easily integrate and compare data sets from different counties and states about the spread of an emerging infectious disease. 


\section{Conclusion}

  Through more efficient management of patient clinical records and patient data at a systems medicine level, SKED could advance patient care towards more predictive and preventative measures that offer the ability to improve individual care, improve overall outcomes, and reduce overall costs associated with patient treatment. We have shown the usefulness of SKED in the interpretation of multi-omic data in clinical disease manifestations and our approach could be extended to general clinical and health management settings.  


\section*{Acknowledgments}
This project was funded in part by Federal funds from the US National Institute of Allergy and Infectious Diseases, National Institutes of Health, Department of Health and Human Services under contract \#HHSN272201200031C, which supports the Malaria Host-Pathogen Interaction Center (MaHPIC).\\

\noindent Special thanks to Sanaz Haghani for the figures. \\

\noindent The author(s) declare(s) that there is no conflict of interest regarding the publication of this paper.


\end{document}